\documentclass[aps,pra,groupedaddress,twocolumn,longbibliography,superscriptaddress]{revtex4}

\usepackage{cancel}
\usepackage{graphicx}
\usepackage{amssymb, amsmath}
\usepackage{bbold}
\usepackage{float}
\usepackage{multirow}
\usepackage{nicefrac}
\usepackage{xcolor}
\usepackage{textcomp}

\newcommand{\diff}{\mathrm{d}}
\newcommand{\out}{{\operatorname{out}}}
\newcommand{\ins}{{\operatorname{in}}}

\newcommand{\cP}{\mathcal{P}}
\newcommand{\cB}{\mathcal{B}}
\newcommand{\cO}{\mathcal{O}}

\newcommand{\mean}[1]{\left\langle #1 \right\rangle}
\newcommand{\means}[1]{\langle #1 \rangle}
\newcommand{\bra}[1]{\langle #1 \vert}
\newcommand{\ket}[1]{\vert #1 \rangle}
\newcommand{\ketbra}[2]{\left\vert #1 \middle\rangle\middle\langle #2 \right\vert}

\usepackage[normalem]{ulem}

\newcommand{\sca}{\nu}

\begin{document}

\title{Superthermal photon bunching in terms of simple probability distributions}

\author{T.~Lettau}
\affiliation{Institut f{\"u}r Physik, Otto-von-Guericke-Universit{\"a}t
  Magdeburg, Postfach 4120, D-39016 Magdeburg, Germany}

\author{H.A.M.~Leymann}
\email{ham.leymann@gmail.com}
\affiliation{Max-Planck-Institut f{\"u}r Physik komplexer Systeme, N{\"o}thnitzer Strasse 38, 01187 Dresden, Germany}
\affiliation{INO-CNR BEC Center and Dipartimento di Fisica, Universita di Trento, I-38123 Povo, Italy}

\author{B.~Melcher}
\affiliation{Institut f{\"u}r Physik, Otto-von-Guericke-Universit{\"a}t
  Magdeburg, Postfach 4120, D-39016 Magdeburg, Germany}

\author{J.~Wiersig}
\affiliation{Institut f{\"u}r Physik, Otto-von-Guericke-Universit{\"a}t
  Magdeburg, Postfach 4120, D-39016 Magdeburg, Germany}

\date{\today}

\begin{abstract}
We analyze the second-order photon autocorrelation function $g^{(2)}$ with respect to the photon probability distribution and discuss the generic features of a distribution that result in superthermal photon bunching ($g^{(2)}>2$).
Superthermal photon bunching has been reported for a number of optical microcavity systems that exhibit processes like superradiance or mode competition.
We show that a superthermal photon number distribution cannot be constructed from the principle of maximum entropy, if only the intensity and the second-order autocorrelation are given. However, for bimodal systems an unbiased superthermal distribution can be constructed from second-order correlations and the intensities alone.
Our findings suggest modeling superthermal single-mode distributions by a mixture of a thermal and a lasing like state and thus reveal a generic mechanism in the photon probability distribution responsible for creating superthermal photon bunching.
We relate our general considerations to a physical system, a (single-emitter) bimodal laser, and show that its statistics can be approximated and understood within our proposed model.
Furthermore the excellent agreement of the statistics of the bimodal laser and our model reveal that the bimodal laser is an ideal source of bunched photons, in the sense that it can generate statistics that contain no other features but the superthermal bunching.
\end{abstract}

\maketitle

\section{Introduction}
\label{sec:introduction}

The second-order photon autocorrelation function $g^{(2)}$ is an important quantity to analyze the statistical properties of a light source \cite{glauber_coherent_1963}.
It can be interpreted as a measure for the coincidence rate of photons and is defined as
\begin{align}
g^{(2)}=\frac{\mean{n^2-n}}{\mean{n}^2},
\end{align}
with the photon number operator $n=b^{\dagger}b$ and it can be measured, e.g., in a Hanbury Brown and Twiss setup \cite{brown_correlation_1956}.

Especially the characterization of light emitted by optical microcavity devices requires the study of the statistical features of the light like $g^{(2)}$, to demarcate various regimes of emission.
For single-photon sources, a value of $g^{(2)}$ well below $0.5$ indicates the creation of a single photon \cite{Michler2000,yuan_electrically_2002}.
In general, values of $g^{(2)}<1$ cannot occur for a classical continuous field, but only for quantized field excitations \cite{garrison_quantum_2014}.

On the other hand, quantum light sources that emit a large number of photons, also require a characterization by $g^{(2)}$ measurements.
The threshold in lasers is indicated by a transition from $g^{(2)}(0)=2$ (typical for thermal states) below the threshold to $g^{(2)}(0)=1$ (typical for coherent/lasing states) above the threshold \cite{wiersig_direct_2009}.
In microlasers the high ratio of spontaneous emission into the lasing mode ($\beta$-factor close to 1) leads to an almost linear behavior of the input-output curve at the threshold and thus hinders the determination of the laser threshold by the intensities alone \cite{rice_photon_1994}.
There are several other indicators of lasing in a microlaser that go beyond the input-output curve like first-order coherence \cite{ates_influence_2008} or leakage into non-lasing modes \cite{musial_correlations_2015}.
However, the change in the photon autocorrelation at the lasing threshold is directly related to the change in the emission mechanism from spontaneous to stimulated emission \cite{loudon_quantum_2000}.
Therefore $g^{(2)}$ is one of the most reliable measures for lasing in microcavity devices \cite{strauf_self-tuned_2006,ulrich_photon_2007,chow_emission_2014}.

When effects become relevant that go beyond spontaneous and stimulated emission into a single cavity-mode from an ensemble of independent emitters, the statistics of the emitted light becomes more intricate.
A very prominent representative for this are the superthermal values of the photon autocorrelation ($g^{(2)}(0)>2$), which will be in the focus of this paper.
Superradiant coupling of the emitters in the gain medium has been reported to lead to $g^{(2)}$-values far above the thermal value \cite{leymann_sub-_2015,jahnke_giant_2016,protsenko_collective_2017,bhatti_superbunching_2015}.
Also the phase difference of coherent laser driving can increase $g^{(2)}$ above $2$ \cite{ciornea_phase_2016}.
Another source that can produce superthermal light is the cathodoluminescence of an ensemble of nitrogen vacancy centers in nanodiamonds \cite{meuret_photon_2015} or the resonance fluorescence of quantum dot-metal nanoparticles \cite{ridolfo_quantum_2010}.
In bimodal lasers, the gain competition
\cite{leymann_intensity_2013,leymann_strong_2013}, dissipative mode coupling
\cite{fanaei_effect_2016}, temporal mode-switching
\cite{redlich_mode-switching_2016}, intermode kinetics
\cite{leymann_pump-power-driven_2017}, external feedback
\cite{hopfmann_nonlinear_2013}, mode coupling \cite{marconi_asymmetric_2016,marconi_stimulated_2016} and a short-pump-pulse-induced quench \cite{marconi_far--equilibrium_2018,javaloyes_superthermal_2017} can lead to superthermal photon autocorrelations. Besides these quantum effects, which are known to produce superthermal photon bunching, there are also pseudo thermal light sources
\cite{jechow_enhanced_2013,assmann_coherence_2012,kazimierczuk_photon-statistics_2015}, which emit intense light with $g^{(2)}=2$ or even exceeding this value \cite{bai_photon_2017,zhou_superbunching_2017}.

Photon correlations have been used since the seminal work of Hanbury Brown and Twiss \cite{hanbury_brown_test_1956}.
It is reported that the large intensity fluctuations present in thermal light can improve the phase sensitivity in interferometry experiments \cite{rafsanjani_quantum-enhanced_2017}, help to detect sub wavelength interference \cite{zhai_direct_nodate}, and improve the reconstruction of photon number distributions by using thermal light as a probe \cite{harder_tomography_2014}.
A high probability of photon pairs, which is indicated by a large $g^{(2)}$, is relevant for applications relying on nonlinear optical processes
\cite{qu_photon_1992,spasibko_multiphoton_2017} like two-photon luminescence microscopy \cite{jechow_enhanced_2013} or thermal ghost imaging \cite{bennink_two-photon_2002,gatti_ghost_2004,kazimierczuk_photon-statistics_2015}.
The aforementioned applications could profit from superthermal photon correlations discussed here, especially when they are created by a bimodal laser where it is known that also the mode that exhibits superthermal bunching has narrow linewidths typical for lasers \cite{khanbekyan_unconventional_2015}.


The photon autocorrelation can be rewritten as  $g^{(2)}=1 + (\textnormal{Var}(n) - \langle n\rangle)/\langle n\rangle^2$ emphasizing that it corresponds to information about the variance of the photon number distribution $P_n$.
To further characterize the statistics of a light source, one can determine higher-order correlations $g^{(k)}$, which contain information about the skewness ($k=3$), the kurtosis ($k=4$) etc.~of the photon number distribution $P_n$.
They can be determined experimentally
\cite{asmann_higher-order_2009,avenhaus_accessing_2010,stevens_high-order_2010,rundquist_nonclassical_2014} and theoretically e.g.~by a cumulant expansion
\cite{foerster_computer-aided_2017,leymann_expectation_2014} or by a direct solutions of the von Neumann-Lindblad equation \cite{gies_theory_2017} (App.~\ref{app:detecton} discusses the problem how the photon distribution and its statistical features like the $g^{(n)}$ inside a leaky cavity transfers to the photon detection statistics outside the cavity).
However, the knowledge of the intensity and first moments of the photon distribution reveals only a fraction of the information contained in the full photon distribution.
For well-known or elementary systems, this information may be sufficient to properly characterize its states.
For more complex or less studied systems, knowledge of the first moments of photon distribution $P_n$ might not be sufficient, since the same value of the autocorrelation can be associated with very different photon statistics.
Recently, direct methods to measure the full photon statistics have been applied to Vertical Cavity Surface Emitting Lasers \cite{wang_nontrivial_2017} and using a transition edge sensor \cite{schlottmann_exploring_2017} or a streak camera \cite{wiersig_direct_2009,asmann_measuring_2010} to microlasers. Using an acousto-optical modulator, it is also possible to generate arbitrary classical photon statistics \cite{straka_generator_2018}.

Understanding which features of the measured statistics are relevant to produce the observed photon bunching effects in general, will help to interpret these experiments.
It is therefore important to discuss the generic features a photon distribution needs to have in order to produce superthermal $g^{(2)}$ values.


The paper is organized as follows: In Sec.~\ref{sec:maxentropie} we employ the maximum entropy method (MEM) to find the simplest unbiased photon distribution that has a superthermal $g^{(2)}$ value.
We demonstrate that only two anticorrelated photon modes can produce such an unbiased photon distribution that contains only information about the intensities and the second-order correlations.
Going a step further in simplification in Sec.~\ref{ssec:fittingmodel}, we then introduce a fitting model for the single-mode distribution based on a linear combination of a low intensity thermal state and a lasing-like state.
In the last part of this section, we discuss the implications of statistics that are composed of incoherent mixtures of simple known states, and demonstrate that these statistics can produce arbitrary high $g^{(2)}$.
In a last step we show in Sec.~\ref{sec:singleemitter} that the introduced fitting model is sufficient to reproduce and interpret the statistics of real physical systems. To this end, we solve the von Neumann-Lindblad equation of a single-emitter bimodal laser and compare its photon statistics to our fitting model.
Section~\ref{sec:concl} concludes the paper. In the appendix \ref{app:detecton} we discuss the detection of the statistical features of a photon distribution from a leaky cavity.

\section{Simplest shape of superthermal distributions}

\subsection{Maximum-Entropy-Method}
\label{sec:maxentropie}
In this section, we discuss which shapes an unbiased photon distribution that produces superthermal $g^{(2)}$ can have.
The standard procedure to create an unbiased distribution $P_n$ from any given information contained in the expectation values $\mean{A_i}= \sum P_n A_i(n)$ is the MEM.
This method creates the maximum entropy distribution (MED), which maximizes the entropy $S = -\sum P_n \ln{P_n}$ under the constraints given by the expectation values $\mean{A_i}$ \cite{fick_quantum_1990}.
Equivalent to maximizing the entropy is finding the Lagrange multipliers $\lambda_i$ of
\begin{align}
  \label{eq:maxentropy}
  P_n =\exp\left({-\sum_{i} \lambda_i A_i(n)}\right)
\end{align}
so that the distribution can be normalized and produces the requested expectation values $\mean{A_i}$.
For a given intensity and photon autocorrelation $\mean{n}, g^{(2)} = f(\mean{n}, \mean{n^2})$ one has to determine three Lagrange multipliers $\lambda_i$ for the operators $A_i = n^i$ ($i=0,1,2$), since the normalization is always implemented by $A_0$.
This can be done by solving the system of equations
\begin{align}
  \label{eq:systemmem}
  \mean{n^j} =\sum_n n^j\exp\left({-\sum_{i=0}^\cO \lambda_i n^i}\right)
\end{align}
for all $\lambda_j$ ($j=0,1,2$) and the $\mean{n^j}$ corresponding to given
intensity and photon autocorrelation. Here, $\cO$ refers to the order of the MED (in our case $\cO=2$).

\begin{figure}
  \includegraphics[width=1\columnwidth]{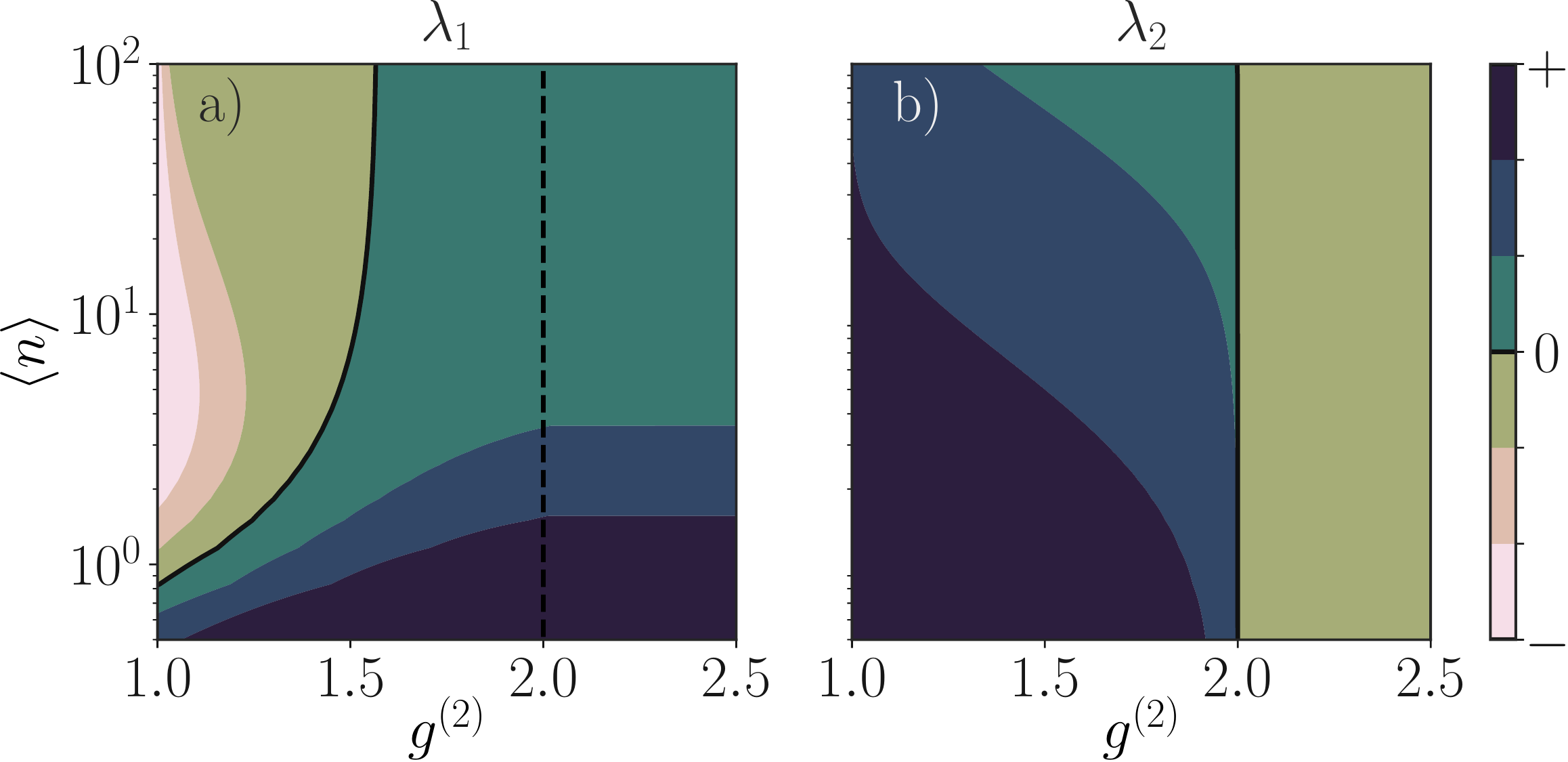}
  \caption{Numerically determined Lagrange multipliers $\lambda_i$ of the second-order MED for a given Intensity $\langle n\rangle$ and autocorrelation $g^{(2)}$. Negative and positive regions are separated by a black curve.
   Autocorrelations above 2 yield negative $\lambda_2$ indicating that the MED cannot be normalized in this case.}
  \label{fig:memlagrange}
\end{figure}

The numerically determined Lagrange multipliers $\lambda_{j}$ are shown in Fig.~\ref{fig:memlagrange} for a wide range of $\mean{n}, g^{(2)}$.
One can see that $\lambda_2$ is only positive for $g^{(2)}<2$ and thus that the second-order MED with an arbitrary large photon number cannot be normalized for any superthermal value of $g^{(2)}$.
The existing values of $\lambda_2$ for $g^{(2)}>2$ are an artifact of the finite number of states used to determine the MED numerically and depend on the number of considered photon states.
In Appendix~\ref{app:upperbound}, we prove analytically that no MED of second-order that has a $g^{(2)}>2$ and an infinite number of photon states exists \cite{einbu_existence_1977}.
Although it is possible to find a MED with \mbox{$g^{(2)}>2$} with a finite number of photon states \cite{dowson_maximum-entropy_1973}, limiting the number of photon states to a maximum value is neither unbiased, nor is it a physically meaningful result.
Note that there is also a lower bound for the photon autocorrelation $g^{(2)} \ge 1 - \nicefrac{1}{\mean{n}}$, which results from the quantized nature of the field \cite{garrison_quantum_2014}.

Since we cannot find a superthermal distribution solely form the knowledge of $(\mean{n}, g^{(2)})$, more information is needed.
Effects that can produce superthermal $g^{(2)}$, like superradiance or mode competition in bimodal lasers, have in common that an additional constituent of the system is correlated with the superthermal photon mode.
This suggests going to a complex system, with additional degrees of freedom, to create a distribution with superthermal $g^{(2)}$.
An alternative way would be to include $g^{(3)}$, i.e.~going to a third-order MED, which results in superthermal distributions that can be normalized.
However, this approach leaves arbitrariness in the much less accessible third-order photon correlation $g^{(3)}$, and leads to distributions that are qualitatively identical to the bimodal ones we discuss below (see App.~\ref{app:3rdmed}), and provides very little insight into the physics of superthermal photon bunching.

\begin{figure}
  \includegraphics[width=\columnwidth]{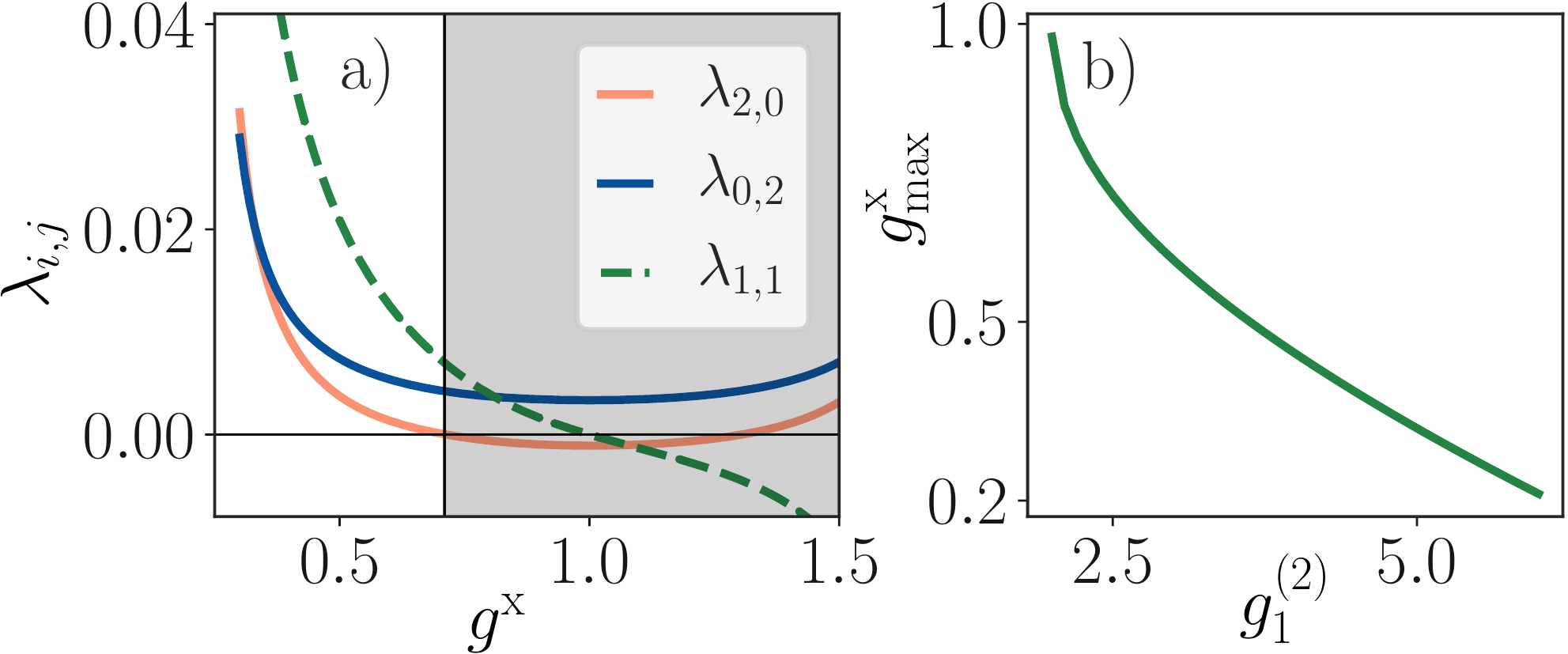}
  \caption{(a): Numerically determined Lagrange multipliers $\lambda_{ij}$ for the bimodal MED of second order, with one mode exhibiting superthermal $g^{(2)}_{1}$ in dependence of the crosscorrelation $g^{\mathrm{x}}$. Only for sufficiently anticorrelated modes all Lagrange multipliers of second-order are positive and thus the MED exists;
  (b): The maximal value of the crosscorrelation $g^{\mathrm{x}}_{\mathrm{max}}$ for a given superthermal $g^{(2)}_1$ in one mode that results in positive $\lambda_{ij}$ of second order. The higher the superthermal $g^{(2)}_{1}$ is, the stronger the two modes have to be anticorrelated.
  Parameters: $\mean{n_{1}}=7$, $\mean{n_{2}}=17$, $g^{(2)}_1 = 2.5$, $g^{(2)}_2 = 1.3$, 80 basis states in each mode.}
  \label{fig:lambdamax}
\end{figure}

\paragraph*{A bimodal system} is the simplest system that allows to derive a second-order MED with superthermal $g^{(2)}$. The general form of the MED of $\cO$th order for a bimodal system reads
\begin{align}
  \label{eq:MEDbimodal}
  P_{n_1,n_2} =\exp\left({-\sum_{i,j=0}^{i+j=\cO} \lambda_{i,j} n_1^i n_2^j}\right).
\end{align}
From $P_{n_1,n_2}$ one can extract the single-mode distribution by summation over the extra degree of freedom, e.g.
\begin{align}
  \label{eq:singlefrombimodal}
  P_{n_1}=\sum_{n_2}P_{n_1,n_2}.
\end{align}
For the bimodal MED of second order, we not only require information about the  individual intensities $\mean{n_i}$ and photon autocorrelations $g^{(2)}_i$ of modes $i=1,2$ but also information about the crosscorrelation
\begin{align}
g^{\mathrm{x}}=\frac{\mean{n_1n_2}}{\mean{n_1}\mean{n_2}}.
\end{align}
Without $g^{\mathrm{x}}$ (i.e. $\lambda_{1,1}=0$), we see that the second-order MED factorizes into a product of two single-mode MEDs.
Since we have already proven, that no superthermal $g^{(2)}$ exist for a single-mode MED of second order, we know that the crosscorrelation of a MED with superthermal photon bunching has to have a non trivial value $g^{\mathrm{x}}\neq 1$.
Figure~\ref{fig:lambdamax}(a) depicts the Lagrange multipliers $\lambda_{2,0}, \lambda_{0,2}$ and $\lambda_{1,1}$, in dependence of $g^{\mathrm{x}}$ for a generic MED with $g^{(2)}_1=2.5$ and $g^{(2)}_2=1.3$. The depicted $\lambda_{i,j}$ need to be positive when an infinite number of photon states is considered. We see that only for sufficiently anticorrelated modes this requirement is fulfilled.
Figure \ref{fig:lambdamax} (b) demonstrates that this observation can be generalized to all second-order bimodal MEDs.
It shows the maximum value of $g^{\mathrm{x}}$ a second-order bimodal MED can have for increasing $g^{(2)}$ values in one of the modes, with the constraint that the MED is normalizable (i.e. $\lambda_{2,0}, \lambda_{0,2}, \lambda_{1,1}>0$).
This reveals that the higher the superthermal photon bunching is the stronger the anticorrelations of the modes need to be.
Figure~\ref{fig:dist_g2_max_entropie} (a) shows a typical bimodal MED with superthermal $g^{(2)}$ and pronounced anticorrelation, visible in the low probability along the $n_1=n_2$ line. Figure~\ref{fig:dist_g2_max_entropie} (b) shows the corresponding single-mode statistics which can be obtained by summing over the other mode as indicated by the $\Sigma$ in panel (a) (more details in the next section).

In summary, two anticorrelated photon modes are in the sense of the MEM the simplest system that can produce superthermal photon bunching, when no information beyond the second-order correlations is to be included.

\subsection{Fitting the single-mode statistics with a mixture of a lasing and a thermal state}
\label{ssec:fittingmodel}

\begin{figure}
  \includegraphics[width=1\columnwidth]{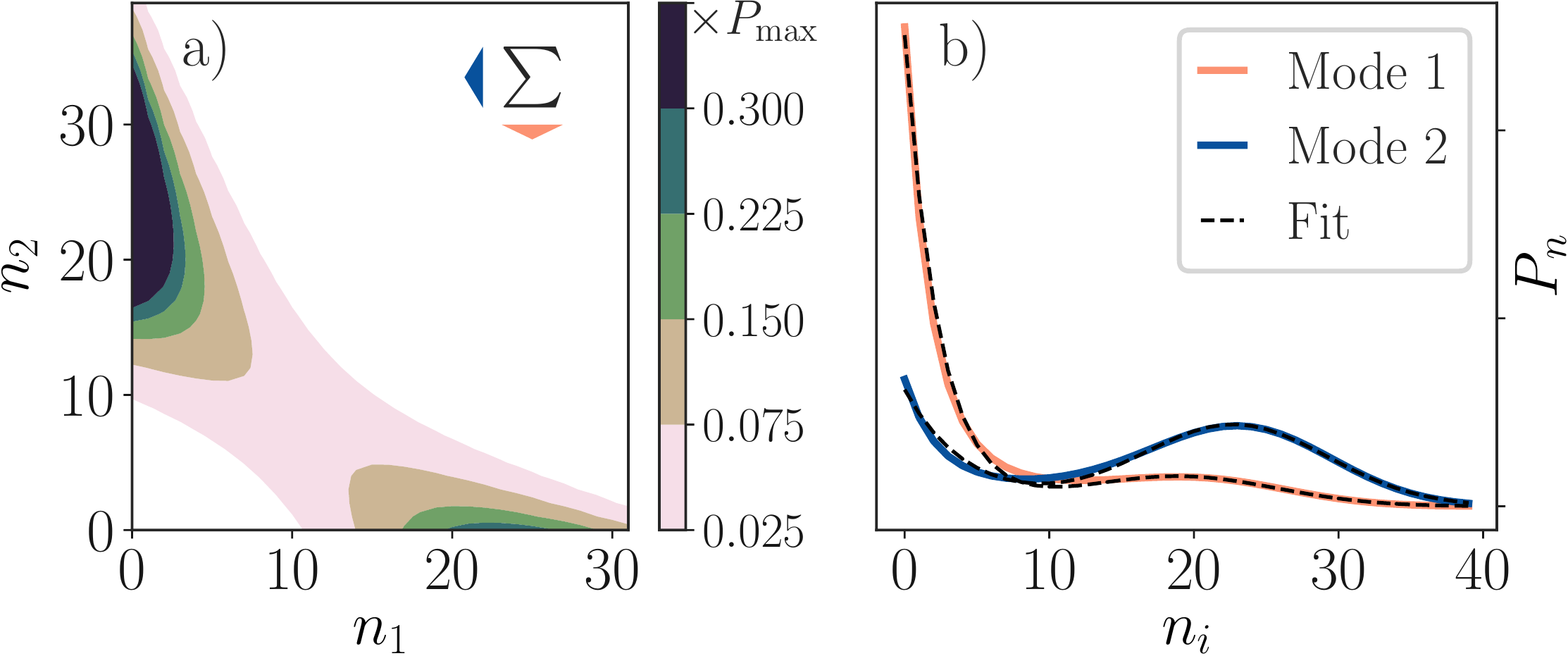}
  \caption{(a): Bimodal MED $P_{n_1,n_2}$ as contour plot; (b): single-mode photon statistics obtained from $P_{n_1,n_2}$ compared to the fitting model introduced in Eq.(\ref{eq:normptherm}). Lagrange multipliers: $\lambda_{1,0} = 0.73$, $\lambda_{0,1}= 0.76$, $\lambda_{2,0}= -0.016$, $\lambda_{0,2}= -0.05$, $\lambda_{1,1}= -0.015$}
  \label{fig:dist_g2_max_entropie}
\end{figure}

A bimodal MED with superthermal $g^{(2)}$ in one of the modes results in a very specific single-mode statistics (see Fig.~\ref{fig:dist_g2_max_entropie} (b)).
The shape of the statistics suggests a fitting model consisting of a linear combination of a thermal distribution $P_n^{\mathrm{T}}$ with a low intensity and a normal distribution $P_n^{\mathrm{N}}$ with an intensity comparable to the one of the original statistics
\begin{align}
  \label{eq:normptherm}
  P_n(\mu,\sigma,\beta,a) &= a\cdot P_n^{\mathrm{N}}(\mu,\sigma) +
    (1-a)\cdot P_n^{\mathrm{T}}(\beta),\\
  P_n^{\mathrm{N}}(\mu,\sigma) &=
    C_N\exp\left(-\frac{(n-\mu)^2}{2\sigma^2} \right),\nonumber\\
     P_n^{\mathrm{T}}(\beta) &=
    C_T\exp\left(-\beta n \right),
    \nonumber
\end{align}
where $\sigma$ is the variance, $\mu$ the center, $C_N$ the normalization constant of the normal distribution and $C_T$ the normalization constant and $\beta$ the effective temperature of the thermal distribution.
Since $n$ is discrete and non-negative, the standard expressions for $C_N$, $\mu$ and $\sigma$ of $P_n^{\mathrm{N}}$, known from the continuous case do not hold.
For large photon numbers (for our purposes photon numbers above $\sim5$) the normal distribution with mean value $\mean{n_{N}}$ and variance $\sigma^2=\mean{n_{N}}$ is a good approximation for a Poisson distribution, which is typical for a lasing state.
The width of the fitted normal distribution is in general larger than the one of the Poisson distribution, which results from the diagonal orientation of the bimodal statistics.
As one can see in Fig.~\ref{fig:dist_g2_max_entropie} (b) this model (depicted by the dashed curve) approximates the single-mode distributions obtained from the second-order MEM very accurately (for details on the fitting procedure see App.~\ref{app:fittsingle}).
The fitting model proposed in Eq.~(\ref{eq:normptherm}) also corresponds well to the notion of two anticorrelated lasing modes, meaning that each mode is not lasing when the other one is, hence $P_n$ has one maximum at $n=0$, and one at $n\approx\mean{n_{N}}$.
The characteristic structure with two maxima, which is reproduced by our fitting model, is well known in the literature and has been observed among others in ring lasers \cite{m-tehrani_intensity_1978,roy_first-passage-time_1980,lett_macroscopic_1981} and QD microlasers \cite{leymann_intensity_2013,leymann_pump-power-driven_2017}.
More importantly, the fitting model reveals a simple mechanism to create superthermal photon bunching in a single-mode, i.e.,~the mixture of a thermal and lasing-like state, created, e.g.,~by switching processes in the time domain \cite{redlich_mode-switching_2016} (see App.~\ref{app:montevonzott}) or a bistability in the switch-on behavior \cite{schlottmann_exploring_2017}.

\begin{figure}[]
	\centering
    \includegraphics[width=1\columnwidth]{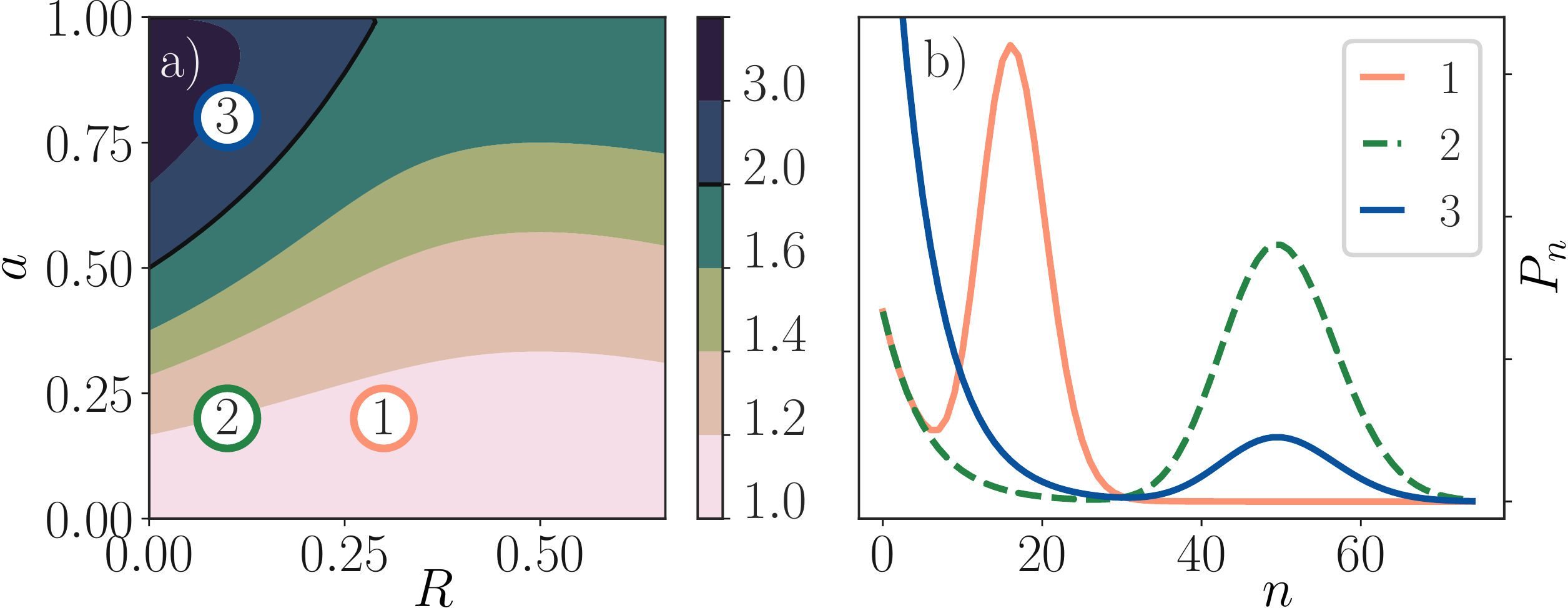}
  \caption{(a): Autocorrelation of a convex combination of two distributions with autocorrelation values $G^1=2$ and $G^2=1$ as a function of the mixing parameter $a$ and the ratio of the intensities $R$. The black curve separates the regions with $g^{(2)}$ values greater and smaller than 2. The values in the darkest area in the upper left corner can become arbitrarily high for small $R$. (b): Single-mode photon statistics for the parameters marked by circles in panel (a).}
  \label{fig:g2_a_R}
\end{figure}

To emphasize the consequences of this model, we show how one can generate arbitrary large $g^{(2)}$ values with a photon distribution
\begin{align}
  P_n = aP_n^1 + (1-a)P_n^2
\end{align}
that is an incoherent mixture of two distributions with known values for $g^{(2)}$ ($G_1$, $G_2$) and $\mean{n}$ ($I_1$, $I_2$).
We choose the indices so that $I_1 \le I_2$ and define the ratio of the intensities as $R = \nicefrac{I_1}{I_2}$ .
The autocorrelation of the composed distribution
\begin{align}
  g^{(2)} = \frac{a G_1 R^2 + (1-a)G_2}{(aR + (1-a))^2}
  \label{eq:g2comp}
\end{align}
depends solely on the ratio of the intensities $R$, the $g^{(2)}$ values of the constituents, and the mixing parameter $a$. Note that this equation was also derived in \cite{grunwald_what_2017} in the context of photon anti-bunching.
The dependence of the resulting $g^{(2)}$ on $a$ and $R$ is shown in Fig.~\ref{fig:g2_a_R} (a) for $G_1=2$ and $G_2=1$, which resembles a composition of a thermal and Poisson distribution.
The black curve marks the parameter region for which $g^{(2)}>2$. Figure~\ref{fig:g2_a_R} (b) shows three examples for a mixture of a thermal and a Poisson distribution with increasing and ultimately superthermal $g^{(2)}>2$.
From Eq.~(\ref{eq:g2comp}) and Fig.~\ref{fig:g2_a_R} (a) we see that a mixture of a thermal and a lasing distribution can create all values of $g^{(2)}\ge 1$. For small $a$ the autocorrelation is almost independent of $R$ and is mainly determined by $a$.
Although $R$ is independent of the absolute value of the intensities, high values of $g^{(2)}$ clearly favor $I^1\approx 0$, especially in microcavity devices where intensities are relatively low.
Note that another consequence of Eq.~(\ref{eq:g2comp}) is that any combination of two statistics with properly chosen $(R,a)$ can produce $g^{(2)}>2$, e.g.~two Poisson or two thermal distributions where the higher temperature distribution acts as the heavy tail of the lower temperature distribution \cite{marconi_far--equilibrium_2018}.

\section{Single-emitter bimodal microcavity laser}
\label{sec:singleemitter}
\begin{figure}[]
	\centering
    \includegraphics[width=1\columnwidth]{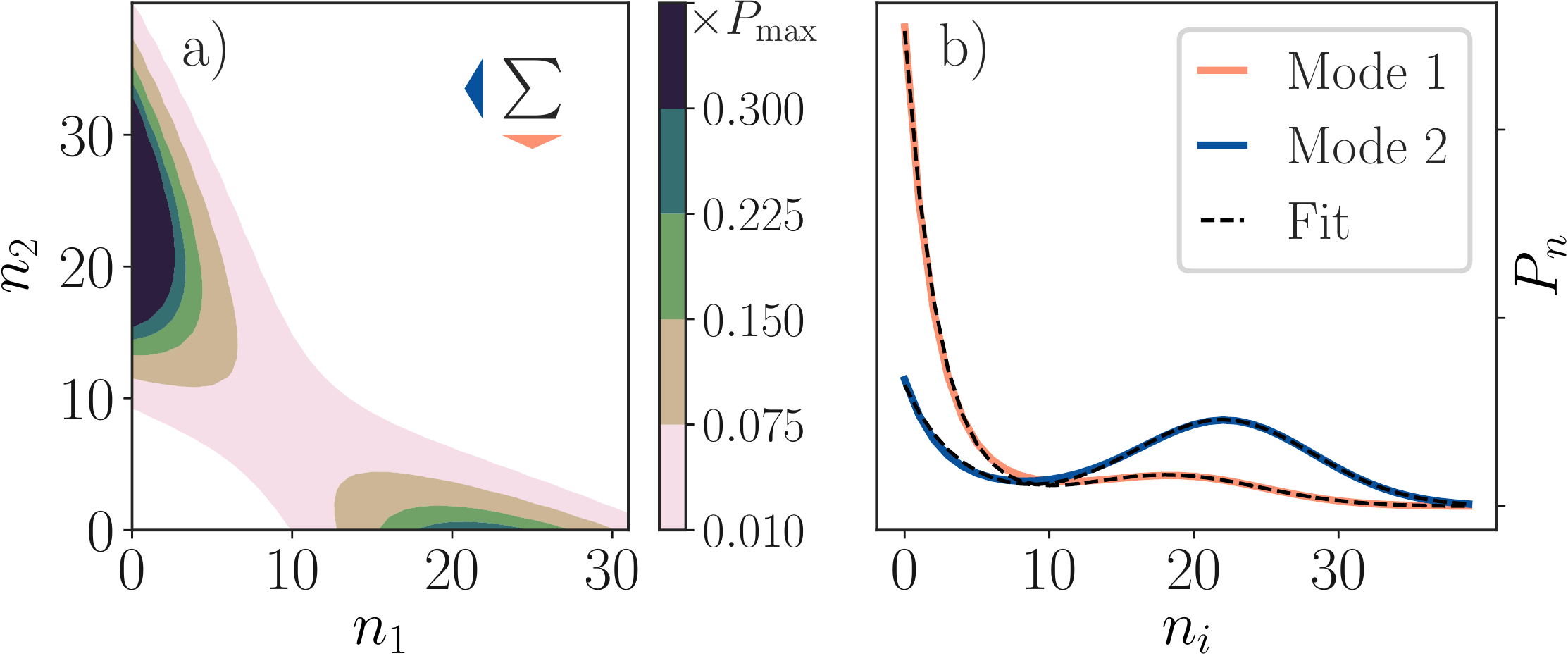}
  \caption{(a): Full photon statistics $P_{n_1,n_2}$ for the single-emitter bimodal laser for pump rate $P = 9.3\sca$; (b): Single-mode photon statistics obtained from $P_{n_1,n_2}$ compared to the fitting model introduced in Eq.~(\ref{eq:normptherm}). The full set of parameters is given in the caption of Fig.~\ref{fig:fittingquali}, where the input-output curves are shown.}
  \label{fig:singleemitterstat}
\end{figure}

In this section, we relate our general considerations and the introduced fitting model to a simple physical model. From a theoretical point of view the simplest laser is a single-emitter single-mode laser \cite{xie_influence_2007,reitzenstein_single_2008,ritter_emission_2010}, so we generalize this to a bimodal laser with a single emitter.
Its steady state is described by the stationary solution of the von Neumann-Lindblad equation
\begin{align}
	\diff_t \rho = &i[H,\rho] + \sum_i \gamma_i \left(L_i \rho L_i^\dagger -\frac{1}{2} L_i^\dagger L_i \rho -\frac{1}{2} \rho L_i^\dagger L_i\right)\label{eq:lindblad}
\end{align}
for the density operator $\rho$.
The Hamiltonian of the system is given by
\begin{align}
  H=&\sum_{i=1,2}\varepsilon_i \ketbra{i}{i}+\sum_{j=1,2}\omega_j b_j^{\dagger}b_j\nonumber\\
  +&\sum_{j=1,2}\left( g_jb_j^{\dagger}\ketbra{1}{2}+g^*_jb_j\ketbra{2}{1}\right),
  \label{Eq:hamiltonian}
\end{align}
where the $\ket{i}$ denote the states of the two-level emitter with energies $\varepsilon_i$ and the $b_j^{(\dagger)}$ the bosonic annihilation (creation) operator for photons in mode $j$ with energy $\omega_j$.
The strength of the light-matter interaction in dipole and rotating wave approximation is given by $g_j$.
The collapse operators $L_i$ and the corresponding rates $\gamma_i$ in the second term in Eq.~\eqref{eq:lindblad} describe the pumping ($L_1=\ketbra{2}{1}$, $\gamma_1=P$), spontaneous emission into non lasing modes ($\ketbra{1}{2}$ , $\gamma_2=\tau_{\mathrm{sp}}^{-1}$), and cavity losses ($L_{3,4}=b_{1,2}$, $\gamma_{3,4}=\kappa_{1,2}$).
As one can see in Eqs.~(\ref{eq:lindblad}) and (\ref{Eq:hamiltonian}), all parameters enter linearly and hence can be scaled by an universal constant $\sca$ which only alters the time scale.
The steady state of Eq.~(\ref{eq:lindblad}) is obtained by numerically integrating the equation with a modified version of QuTip \cite{johansson_qutip_2013}.
With the resulting density operator we can compute the two-mode photon statistics $P_{n_1,n_2}=\mathrm{Tr}(\ketbra{n_1,n_2}{n_1,n_2}\rho)$, which is depicted in Fig.~\ref{fig:singleemitterstat} (a), and all other desired observables.

Figure \ref{fig:singleemitterstat} has a striking resemblance with Fig.~\ref{fig:dist_g2_max_entropie}, revealing that this single-emitter bimodal laser can generate almost perfectly unbiased superbunching in the sense of the maximum entropy principle.
We also see in Fig.~\ref{fig:singleemitterstat} (b) that the proposed fitting model [Eq.~(\ref{eq:normptherm})] is, in analogy to the results of the previous section, very well suited to approximate and interpret this type single-mode statistics (see App.~\ref{app:fittsingle} for fitting parameters).

\begin{figure}[]
	\centering
    \includegraphics[width=1\columnwidth]{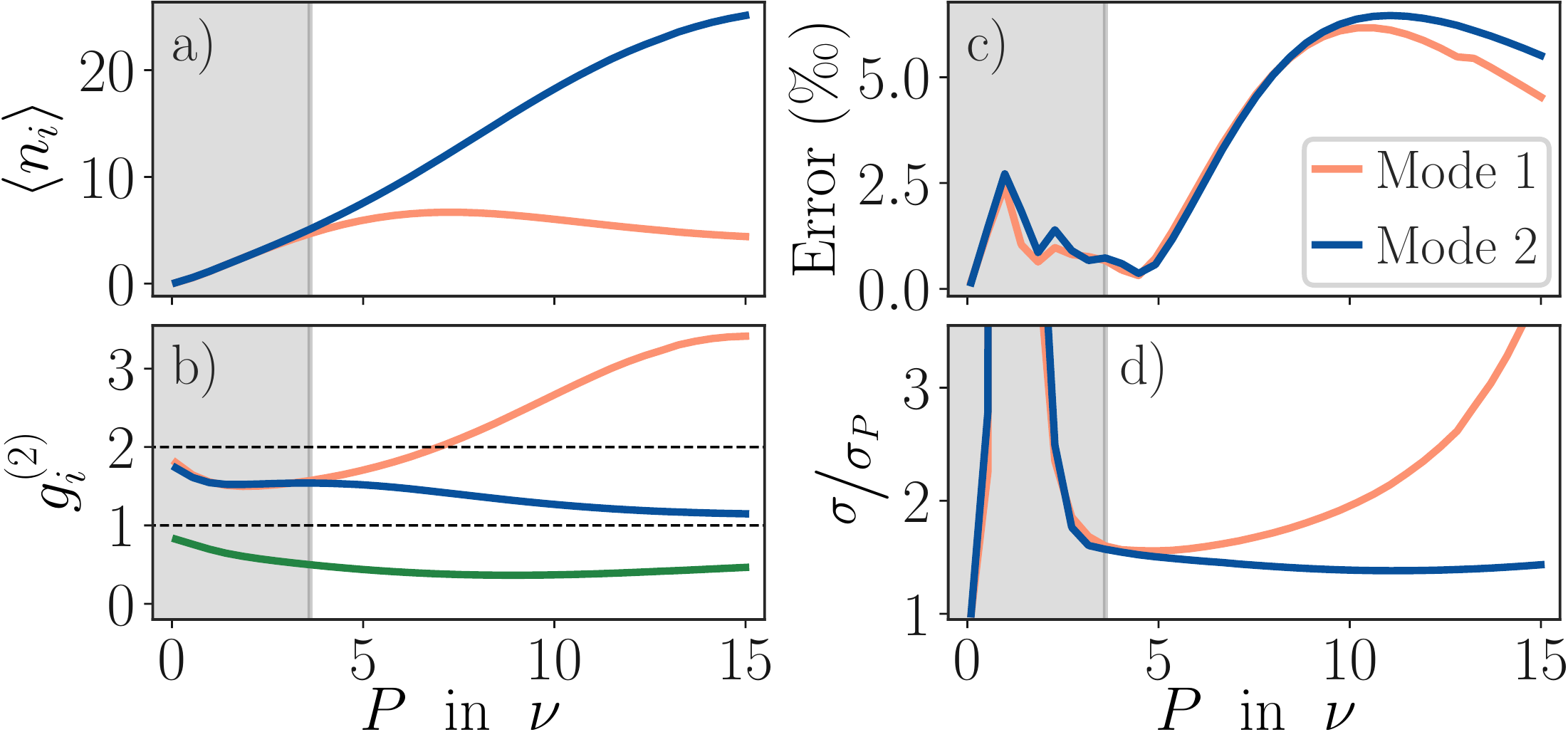}
  \caption{Laser characteristics and characteristics of the fitting model versus the dimensionless pump power. Orange curves belong to mode 1, blue curves to mode 2, the gray area marks the region where the fitting model is not suitable and the fitting routine does not converge without manual aid.
  (a): Input-output characteristics for the intensities $\mean{n_i}$. (b): Photon autocorrelations $g^{(2)}_i$ and the crosscorrelation $g^{\mathrm{x}}$ (depicted in green).
  (c): Root-mean-square deviation of the fitting model in Eq.~(\ref{eq:normptherm}) from the actual single-mode distribution of the bimodal laser in \textperthousand.
  (d): Variance $\sigma$ of the normal distribution in the fitting model divided by the variance $\sigma_P$ of the Poisson distribution with the same mean photon number.
  Parameters: $g_1\!=\!\sca$, $g_2\!=\!0.96\sca$,
  $\omega_1\!=\!0.2\sca$, $\omega_2\!=\!\sca$,
  $\varepsilon_1\!=\!0$, $\varepsilon_2\!=\!\sca$,
  $\tau_{\mathrm{sp}}\!=\!\nicefrac{\sqrt{2}}{\sca}$,
  $\kappa_1\!=\!\nicefrac{0.16}{\tau_{\mathrm{sp}}}$, $\kappa_1\!=\!\nicefrac{0.17}{\tau_{\mathrm{sp}}}$}
  \label{fig:fittingquali}
\end{figure}

Figures~\ref{fig:fittingquali} (a) and (b) depict the intensities $\mean{n_i}$, the photon autocorrelations $g^{(2)}_i$ and the crosscorrelation $g^{\mathrm{x}}$ for increasing pump rates.
We see the typical behavior of a bimodal laser \cite{tehrani_coherence_1978,mandel_optical_1995,leymann_intensity_2013,fanaei_effect_2016}: at the lasing threshold the competition for the limited gain sets in and in this case mode $1$ (orange curves) is loosing while mode $2$ (blue curves) is winning the gain competition.
Furthermore, the losing mode exhibits superthermal photon correlations and the two modes are strongly anticorrelated ($g^{\mathrm{x}}\approx 0.5$ green curve).
For pump rates exceeding $15\sca$ (not shown) we observe the typical quenching effect of a single two-level emitter \cite{jones_photon_1999}.
In Fig.~\ref{fig:fittingquali} (c) the deviation of the fitting model [Eq.(\ref{eq:normptherm})] from the actual single-mode distribution $P_{n_i}$ is depicted. We see that for all pump values the error is significantly less than 1\%, meaning that the deviation of the plotted distributions is barely visible.
For small pump rates, the proposed fitting model does not converge well (gray area) and the error curve behaves quite erratic indicating a certain arbitrariness of the fitting parameters.
Indeed, this gray area marks the pump region in which the proposed fitting model is not appropriate since the maximum of the thermal and the lasing-like state are not yet separable.
The low fitting error in this region is a result of the simple form of the statistics and the small number of $P_{n_i}$ with non zero occupation.
However, this is not a downside of our fitting model since it is not designed to describe the photon statistics for all possible pump powers, but to fit and interpret the photon statistics leading to superthermal photon bunching above the lasing threshold.

To produce the data in Fig.~\ref{fig:fittingquali} we used Eq.~(\ref{eq:normptherm}) for each mode separately, in particular, we have allowed different mixing parameters $a_i$. Since the two single-mode distributions originate from a single bimodal distribution with strong anticorrelations between the modes, the mixing parameters $a_i$ are not independent. Indeed, Fig.~\ref{fig:fitparameter} (c) in App.~\ref{app:fittsingle} clearly shows that for pump powers above the laser threshold the mixing parameters add up to unity. This justifies an ansatz for a bimodal fitting model (Eq.~(\ref{eq:twomodefit})), which relates to the observation that the system is in one of two distinct states; (i): mode
1 lases and mode 2 is thermal; (ii): mode 1 is thermal and
mode 2 lases. In this new ansatz only a single mixing parameter $a$ exists which describes the mixing between state (i) and (ii). In App.~\ref{app:montevonzott} we discuss how this mixing parameter $a$ can be interpreted within the framework of quantum trajectories as an average dwell time in one of these states.

Figure \ref{fig:fittingquali} (d) shows the variance $\sigma$ of the laser-like part of the fitting model compared to the variance $\sigma_P$ of a Poisson distribution with the same intensity.
Above the lasing threshold, the variance of the laser-like part of the winning mode 2 is almost constant and close to the one of the Poisson distribution ($\sigma\approx 1.4\sigma_P$), consistent with its lasing character.
However, the variance of the losing mode 1 is increasing drastically and rises to values that exceed three times the values of the corresponding Poisson distributions.
This shows that the notion of switching between non-lasing (thermal distribution) and lasing (Poisson distribution) is to simplistic to describe this system and that it is rather a switching between a non-lasing and a broadened laser-like state, as described by our fitting model.
Nevertheless we clearly see that the mixture of two simple states, corresponding to the notion of a spontaneous temporal switching, is very helpful to analyze superthermal statistics.

\section{Conclusion}
\label{sec:concl}
We discussed the general features of photon statistics with of superthermal photon bunching.
Using the principle of maximum entropy we have demonstrated that no unbiased single-mode photon statistics with $g^{(2)}>2$ can be constructed without knowledge of its higher moments.
We concluded that two anticorrelated modes are the simplest system which exhibits superthermal $g^{(2)}$ and provides insight into the physics behind superthermal light sources.
In accordance with results obtained from the von Neumann-Lindblad equation, the bimodal maximum entropy distribution justifies a fitting model consisting of a mixture of a low intensity thermal and a high-intensity lasing-like state for the single-mode distributions.
This model reveals a generic mechanism to create arbitrary high $g^{(2)}$, by pushing a small fraction of a lasing like state to large photon numbers in an otherwise thermal state.
The proposed model approximates the statistics of a single-emitter bimodal laser very well.
It is remarkable that this bimodal laser produces the simplest possible superthermal statistics in the sense of the maximum entropy method, revealing that a bimodal laser is an ideal system to generate a superthermal statistics without additional correlations.

\begin{acknowledgments}
T.~Lettau and H.A.M.~Leymann have contributed equally to this work. B.~Melcher acknowledges funding from the DFG (Project No. WI1986/9-1).
\end{acknowledgments}

\appendix

\section{Measurement of higher-order photon autocorrelation functions by detection of leaked photons}
\label{app:detecton}
To be able to interpret the statistical properties of the light field, it is important to know whether they are the same on the inside and on the outside of a light emitting device.
To this end we apply the general results obtained in \cite{lee_external_1993} to the problem of the measurement of the autocorrelation function.

\paragraph{Decaying cavity field:} One elementary model to describe the leakage of a cavity and to transfer the light field outwards was proposed in \cite{lee_external_1993}.
The author assumed that the leakage of photons is the only relevant process, especially that the cavity is not pumped, when the measurement starts at time $t_1$.
The probability $P_l^\out$ to find $l$ leaked photons at time $t_2$ if the cavity contains initially $n$ photons is described by
\begin{align}
  \label{eq:berndiff}
  \frac{\diff}{\diff t_2}P_l^\out = -(n-l)\eta P_l^\out + (n-l+1)\eta P_{l-1}^\out,
\end{align}
where $\eta$ is the loss rate of the cavity mode.
The solution of this equation in terms of the initial distribution inside the cavity $P_n^\ins$ is
\cite{lee_external_1993}
\begin{align}
  \label{eq:bernoulli}
  P_l^\out (t_1, t_2) &= \sum_{n=l}^\infty P_n^\ins {n \choose{l}}
    (1 - \zeta)^{n-l}\zeta^{l} = \cB \cdot P^\ins\\
  \zeta &= \zeta(t_1, t_2) = (e^{-\eta t_1} - e^{-\eta t_2})^l, \nonumber
\end{align}
where $\cB$ is a matrix with binomial distributions in its columns, which means that each initial $P_n^\ins$ is weighted by a binomial distribution of order $n$.

This transformation has an interesting property:
If we connect the autocorrelation functions of arbitrary order
\begin{align}
  \label{AppEq:gk}
  g^{(k)}=\frac{\means{\prod_{i=0}^{k-1}(n-i)}}{\means{n}^k}
\end{align}
from the outside with the ones on the inside, we find a simple relation for the involved expectation values
\begin{align}
  \label{AppEq:mean}
  \mean{\prod_{i=0}^{k-1}(n-i)}_\out &= \zeta^{k}
                                        \mean{\prod_{i=0}^{k-1}(n-i)}_\ins
\end{align}
and therefore that all $g^{(k)}$ are the same on the inside and the outside and do not change in time.

Besides the physical interpretation, Eq.~(\ref{eq:bernoulli}) produces a mapping that allows scaling a distribution with $\mean{n}$ to another distribution that has the same $g^{(k)}$ but a smaller mean value $\mean{\tilde n} \in [0, \mean{n}]$.
Although there is an inverse transformation $\cB^{-1}$\cite{lee_external_1993}, it is not possible to use this transformation to find a distribution with the same statistical features as the original one (same $g^{(k)}$) but with a larger mean value $\mean{\tilde n}$.

\paragraph{Continuously pumped cavity field:} To model the detection of photons leaking out of a continuously pumped cavity, we assume that the cavity is already in a steady state and that every emitted photon is immediately fed back into the cavity by the internal dynamics, or rather that the fluctuations are small compared to the amount of photons.
Under this premises, Eq.~\eqref{eq:berndiff} changes to
\begin{align}
  \label{eq:poissdiff}
  \frac{\diff}{\diff t}P_l^\out = -n\eta P_l^\out + (n+1)\eta P_{l-1}^\out,
\end{align}
i.e., the time derivative of $P_l^\out$ no longer depends on the number of
already leaked photons and it is only necessary to use one time $t$, since
$P_n^\ins$ is in a steady state. This equation has the solution
\begin{align}
  \label{eq:poissonT}
  P_l^\out(t) &= \sum_{n=0}^\infty P_n^\ins
    \frac{(n\eta t)^l}{l!}e^{-n\eta t} = \cP \cdot P^\ins,
\end{align}
where $\cP$ is a matrix with Poisson distributions in its columns, which means that now each initial $P_n^\ins$ is weighted by a Poisson distribution with mean value $n \eta t$.
This time we find that
\begin{align}
  \mean{\prod_{i=0}^{k-1}(n-i)}_\out &= (\eta t)^k \mean{n^k}_\ins
\end{align}
and therefore, that the autocorrelations on the outside
\begin{align}
  g^{(k)}_\out = \frac{\mean{n^k}_\ins}{\mean{n}^k_\ins}
\end{align}
are still constant over time, but not equal to those on the inside.
In the second-order, the autocorrelation is always larger on the outside
\begin{align}
  g^{(2)}_\out = g^{(2)}_\ins + \frac{1}{\mean{n}_\ins} > g^{(2)}_\ins.
\end{align}
However, since this model is only valid for relatively large values of $\mean{n}_\ins$, the difference in the autocorrelation of second order is insignificant.
In contrast to the first transformation, this mapping allows to scale the initial distribution to an arbitrary mean value.

\section{Second and third-order maximum entropy distribution}
\subsection{Poof of the upper bound of $g^{(2)}$ in the second-order maximum entropy distribution}
\label{app:upperbound}

It was shown by \cite{einbu_existence_1977} that every continuous MED of $\cO\!-\! 1$th order implies a boundary for the $\cO$th moment of the $\cO$th-order MED.
All steps of this proof are also valid for discrete distributions.
One can show that the sum over the products of the total differentials of the Lagrange multipliers $\diff\lambda_i$ and the corresponding moments $\diff \means{n^i}$
\begin{align}
  \label{eq:proofineq}
  \sum_{i=0}^\cO \diff\lambda_i \cdot \diff\means{n^i} \le 0
\end{align}
is always smaller than zero, and the equality holds for the trivial case $\diff\lambda_i=0, \forall i$.
To take advantage of this inequality, one considers a valid $\cO - 1$th-order MED with Lagrange multipliers $(\lambda_0, \dots, \lambda_{\cO-1}, \lambda_{\cO}=0)$ and moments $(1, \means{n^1}, \dots, \means{n^\cO})$.
If the $\lambda_i$ in the $\cO$th-order MED are changed in such a way, that only the moment $\means{n^\cO}$ is altered, Eq.~(\ref{eq:proofineq}) simplifies to
\begin{align*}
  \diff\lambda_\cO \cdot \diff\means{n^\cO} \le 0.
\end{align*}
But, in order to normalize the MED, the Lagrange multiplier of highest order has to be positive and therefore we find $\diff\lambda_\cO \ge 0 \Rightarrow \diff\means{n^\cO} \le 0$, i.e., the moment $\means{n^\cO}$ in the $\cO$th order is less or equal than the $\means{n^\cO}$ in the $\cO\!-\! 1$th order, if all other moments stay the same.

In particular, in the first order $(P_n \propto e^{-\lambda_1 n})$, we have the moments $(1, \means{n})$ and can calculate $\means{n^2}=2\means{n}^2 + \means{n}$ analytically. If we go to the second order and keep the moments $(1, \means{n})$, we find immediately, that $\means{n^2} \le 2\means{n}^2 + \means{n}$ and for the autocorrelation
\begin{align}
  g^{(2)} \le \frac{2\means{n}^2 + \means{n} - \means{n}}{\means{n}^2} = 2.
\end{align}
This finding does \emph{not} generalize to higher orders.
We used the fact that we can construct a MED to every positive $\means{n}$ in the first order and therefore we have an upper bound in $g^{(2)}$ for every valid set of Lagrange multipliers in the second order.
But, not every pair of moments $(\means{n}, \means{n^2})$ that can be created in the third-order is also valid in the second order, i.e. there is no general constraint for $g^{(3)}$ in the third-order.

\subsection{Third-order Maximum Entropy Distribution}
\label{app:3rdmed}
\begin{figure}[]
	\centering
    \includegraphics[width=1\columnwidth]{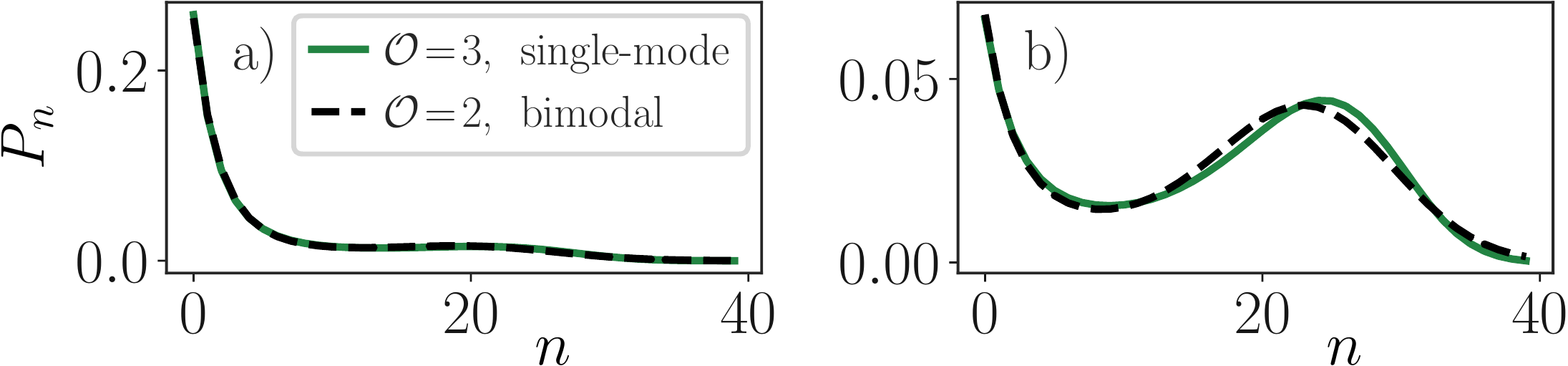}
  \caption{Third-order single-mode MED (solid green curve) compared to a single-mode distribution derived from the corresponding second-order bimodal MED (a) mode 1, (b) mode 2. Lagrange multipliers: $\cO=2$: see caption of Fig.~\ref{fig:dist_g2_max_entropie}; $\cO=3$: $\lambda_1 = 0.37$, $\lambda_1 = -0.028$, $\lambda_3 = 0.00057$.}
  \label{fig:3rdmed}
\end{figure}

Figures~\ref{fig:3rdmed} (a) and (b) show third-order single-mode MEDs, compared to the corresponding single-mode distributions derived from a second-order bimodal MED. The distributions are virtually identical, however in the main text we discuss only the bimodal MED. To construct a third-order MED one needs additional information form $g^{(3)}$, which we do not have at hand, and it introduces additional arbitrariness. However, the main reason for preferring the bimodal distribution over the third-order MED is that the latter does not allow for deeper insight into the physics of superthermal photon bunching.

\section{Fitting the model to the single-emitter bimodal laser}
\label{app:fittsingle}
\begin{figure}[]
	\centering
    \includegraphics[width=1\columnwidth]{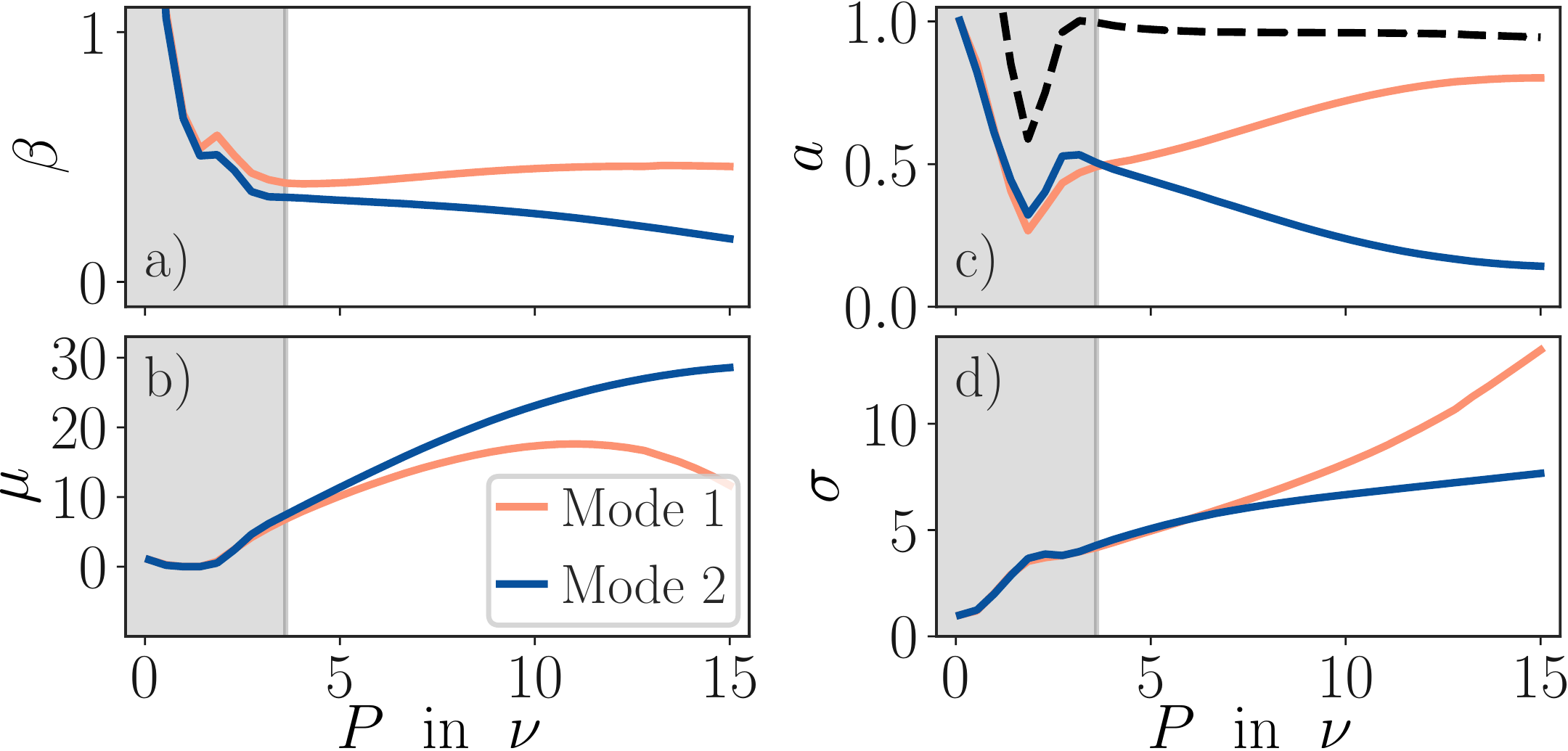}
  \caption{Parameters of the fitting model for Fig.~\ref{fig:fittingquali}; (a): The inverse temperature of the thermal distribution; (b): The center of the normal distribution; (c): The mixing parameter of both modes and their sum depicted as black dashed curve ; (d): The standard deviation, i.e., the width of the normal distribution. The gray shaded area marks in region for which the fitting model is not applicable.}
  \label{fig:fitparameter}
\end{figure}

To fit the four parameters required in Eq.~\eqref{eq:normptherm}, we have minimized the root-mean-square deviation between the fitted and the original distribution with SciPy's implementation of the Broyden-Fletcher-Goldfarb-Shanno algorithm.
The results for the input-output characteristics presented in Fig.~\ref{fig:fittingquali} are shown in Fig.~\ref{fig:fitparameter}.
For small pump rates (gray shaded region), the dependence of the parameters on the pump rate is quite different from the remaining part, which is best seen in panel (c).
Initially, the distribution is in a pure thermal state ($a=1$). In the interval $P=(0,2\sca]$ it is then broadened by the addition of a normal distribution with $\mu\approx 0$. At $P=2\sca$ the parameter $a$ increases abruptly. At this pump value the additional normal distribution, becomes visible by forming a turning point in $P_n$.
In accordance with the strong anticorrelation for higher pump rates $\mu$ and $a$ increase up to the point, where both modes split.

In the gray shaded region, we had to regularize the cost function to exclude negative values of $\mu$ and values of $a$ greater than one and avoid local minima of the cost function manually.
The relatively small errors ($\le 0.25\%$) visible in Fig.~\ref{fig:fittingquali} are predominantly a result of the simple shape and the small number of relevant states of $P_n$.
However, in the remaining part (white area), after the modes have split and taken the shape consistent with the fitting model, the fit routine converges very stable.
Furthermore, the extracted parameters can provide further insight.
(1) The inverse Temperature $\beta$ is given by the logarithm of the slope of the distribution at $n=0$ [see Eq.~(\ref{eq:normptherm})].
(2) Since the thermal part in the composite distribution is scaled by $a$, this parameter can be obtained by dividing $P_0$ of the original distribution by $P_0^{\mathrm{T}}$.
(3) The center of the normal distribution $\mu$ is approximately the mean value and can be estimated in terms of $a$, the mean value of $P_0^{\mathrm{T}}$ and of the one of the original distribution.

As depicted by the dashed curve in Fig.~\ref{fig:fitparameter} (c), the mixing parameters $a$ of both modes add up to one.
This originates from the separation of the two maxima in $P_{n_1,n_2}$ corresponding to the two modes (Fig.~\ref{fig:singleemitterstat}), i.e., from the strong anticorrelation between the modes.
Since lasing in one mode means non lasing in the other mode, the thermal part with weight $a$ of one mode is the lasing part with weight $1\! -\! a$ of the other mode.
This observation justifies the ansatz
\begin{align}
  P_{n_1, n_2} = a\cdot P_{n_1}^{\mathrm{T}}P_{n_2}^{\mathrm{N}}
  + (1-a)\cdot P_{n_2}^{\mathrm{T}}P_{n_1}^{\mathrm{N}}.
  \label{eq:twomodefit}
\end{align}
for the full two-mode statistics, which is the simplest ansatz resulting in the single-mode fitting model in Eq.~(\ref{eq:normptherm}).
This demonstrates that although the information about correlations between the modes is lost in the single-mode distributions, clear traces of the anticorrelation between two distinct states of the system (i) and (ii), as defined in Sec.~\ref{sec:singleemitter}, can still be extracted from the structure of the two single-mode distributions using the proposed fitting model. The anticorrelation between the two different states (i) and (ii) will be further examined in the next section.

\section{Monte-Carlo Trajectories}
\label{app:montevonzott}
\begin{figure}
  \includegraphics[width=1\columnwidth]{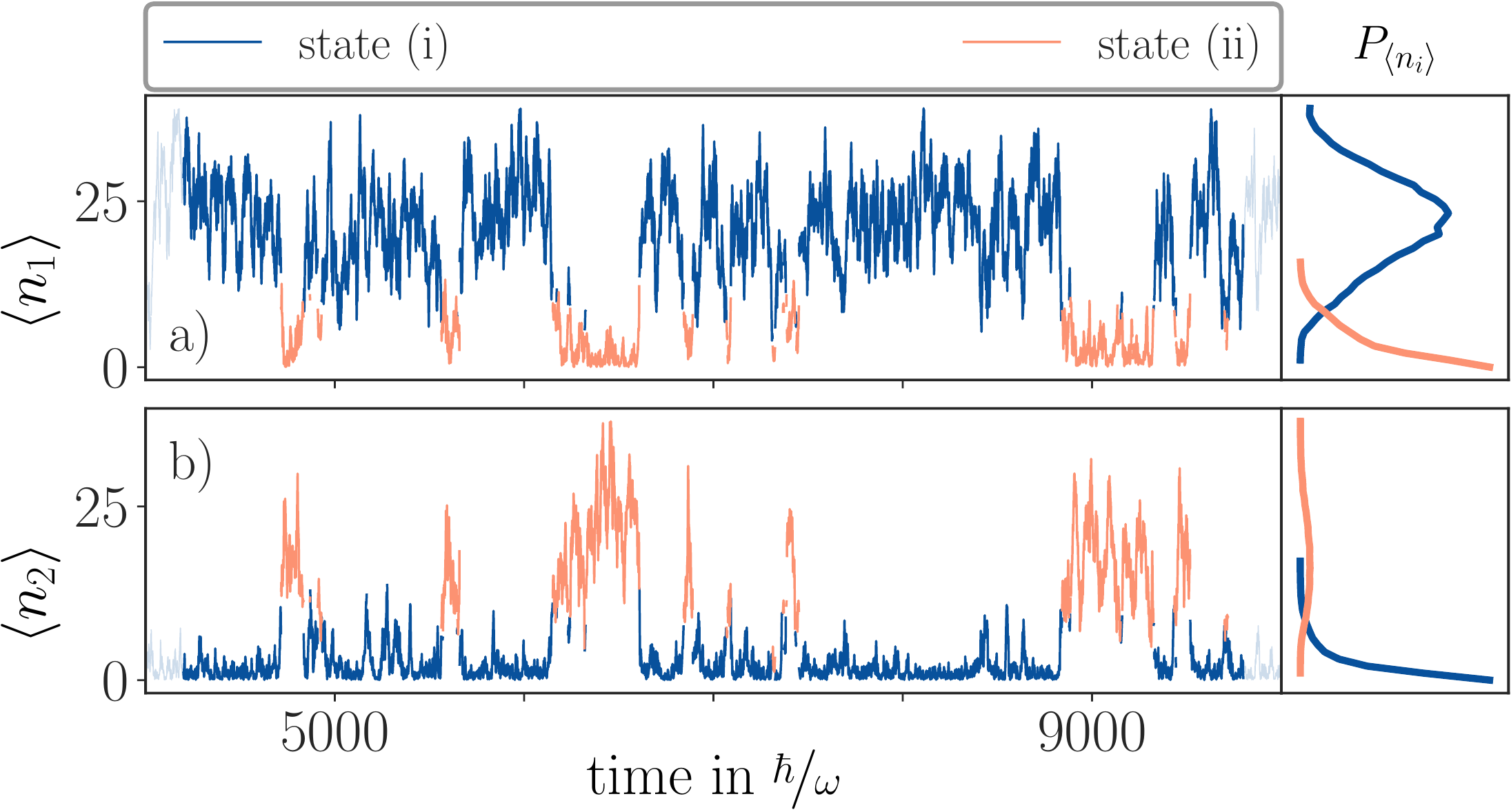}
  \caption{Part of $\bra{\psi^k}n_j \ket{\psi^k}(t)$ from a Monte-Carlo trajectory $\ket{\psi^k}(t)$ calculated from the vNL equation for the bimodal laser with the same parameters as in Fig.~\ref{fig:singleemitterstat}; The values of $\bra{\psi^k}n_j \ket{\psi^k}(t)$ with $n_1 >(<) n_2$ are identified with state (i(ii)), respectively. The probability over time for each mode and state are depicted on the right margin (We used $10^6$ time steps to create the statistics).}
  \label{fig:montecarlo}
\end{figure}

To gain intuition about the dynamics of the system we unravel the von Neuman Lindblad equation [Eq.~(\ref{eq:lindblad})] in an ensemble of quantum trajectories \cite{carmichael_open_1993,breuer_theory_2002}.
For a pumprate well above the lasing threshold, we have depicted $\bra{\psi^k}n_j \ket{\psi^k}(t)$ in Fig.~\ref{fig:montecarlo} for each mode, which results from a part of such trajectory $\ket{\psi^k}(t)$. We associate all occupations for which $\mean{n_1} > \mean{n_2}$ holds, with state (i) (blue), and accordingly $\mean{n_1} < \mean{n_2}$ with state (ii) (orange).
On the right margin of Fig.~\ref{fig:montecarlo} we show the statistics $P_{\mean{n_{j}}}$, build up from a single trajectory $\bra{\psi^k}n_j \ket{\psi^k}(t)$ over time, with a total time of $10^6\nicefrac{ \hbar}{\omega}$ and $10^6$ sample points.
Besides the remaining noise, which would vanish for infinite calculation time, the resulting statistics clearly correspond to those shown in Fig.~\ref{fig:singleemitterstat} and can be separated into a thermal and a normal distributed part weighted with $\tilde{a}$.
In this dynamical picture $\tilde{a}$ can be interpreted as the dwell time fraction in one of the states (i) and (ii), and the system is spontaneously switching between them \cite{lippitz_markus_statistical_2005,jung_current_2002,budini_open_2009}.
When the two parts of the statistics in the fitting model Eq.~(\ref{eq:normptherm}) (thermal and lasing like) are well separated, as in this case, the value of the dwell time parameter $\tilde{a} = 0.32$ and the mixing parameter $a = 0.30$ are almost identical.

\end{document}